\documentclass[runningheads]{llncs}
\usepackage{graphicx}
\usepackage{hyperref}
\begin{document}
\title{Towards the Baikal Open Laboratory in Astroparticle Physics}
%
%
\author{
Pavel Bezyazeekov\inst{1}\and
Igor Bychkov\inst{2}\and
Nikolay Budnev\inst{1}\and
Daria Chernykh\inst{1}\and
Yulia Kazarina\inst{1}\and
Dmitriy Kostunin\inst{3}\and
Alexander Kryukov\inst{4}\and
Roman Monkhoev\inst{1}\and
Alexey Shigarov\inst{2}\and
Dmitriy Shipilov\inst{1}}

\authorrunning{Y. Kazarina et al.}
%
\institute{Applied Physics Institute, Irkutsk State University, Irkutsk, Russia \and
Matrosov Institute for System Dynamics and Control Theory, Siberian Branch of Russian Academy of Sciences, Irkutsk, Russia \and 
DESY, Zeuthen, Germany \and
Lomonosov Moscow State University, Skobeltsyn Institute of Nuclear Physics, Moscow, Russia
}
%
\maketitle              
\begin{abstract}
The open science framework defined in the German-Russian Astroparticle Data Life Cycle Initiative (GRADLCI) has triggered educational and outreach activities at the Irkutsk State University (ISU),
which is actively participated in the two major astroparticle facilities in the region: TAIGA observatory and Baikal-GVD neutrino telescope.
We describe the ideas grew out of this unique environment and propose a new open science laboratory based on education and outreach as well as on the development and testing new methods and techniques for the multimessenger astronomy.

\end{abstract}
\section{Introduction}
The only way to study the high-energy processes occurring outside our Galaxy is to detect the radiation and ultra-high energy particles generated by these processes. 
When colliding with the atmosphere these particles produce secondary cascades, namely extensive air-showers (EAS).
Reaching the surface of the Earth, these cascades can cover areas of tens of square kilometers.
However, with an increase of the primary energy, the flux falls steeply, reaching one particle per year per thousand square kilometers.
It is a main reason why the modern astrophysics is moving towards consolidation and integration of facilities aimed at the detection of various cosmic \emph{messengers}~\cite{Kowalski_Bartos_mm}.

The large-scale astroparticle physics implies the life cycle of experiments in the order of few dozens years, what means the data will be acquired and analyzed by the several generations of the physicists.
Thus, not only the data life cycle has to be properly maintained for the sustainability of experiments, but the human aspects, e.g. training and continuity, have to be taken into account as well.

In this work we continue development of the outreach and educational framework declared in the German-Russian Astroparticle Data Life Cycle Initiative~\cite{Bychkov:2018zre}.
Tightly connected to the Data Life Cycle, this framework requires open data and software policies and aimed at the training of future experts in the astroparticle physics as well as at the outreach of this field.

In our case we have a unique environment, which allows us develop towards establishing of the \textit{Baikal Open Laboratory} in Astroparticle Physics:
\begin{itemize}
\item \textit{International GRADLCI framework} provides an informational support for our activity (e.g. platform \texttt{astroparticle.online}) and increases a visibility of outreach and education activity related to astrophysics. 
\item \textit{Cooperation with astrophysical facilities in Baikal region}, namely with TAIGA (Tunka Advanced Instrument for cosmic rays and Gamma Astronomy) observatory~\cite{Budnev:2018fxf} and Baikal-GVD (GigaVolume Detector) neutrino telescope~\cite{Avrorin:2013uyc} helps us to stay connected with high-level experimental astrophysics.
Moreover, the historical, geographical and infrastructure connections between these experiments and members of GRADLCI enhance the integration of data life cycle and open data policies into operating experiments and gives unique options for testing of these policies.
\item \textit{Educational resources at the Irkutsk State University (ISU)}. 
Besides bachelor and master programs in particle and astroparticle physics, ISU organizes
two famous international schools in this field, namely Baikal Young Scientists' International School on Fundamental Physics\footnote{\url{http://bsfp.iszf.irk.ru/}} and 
Baikal Summer School on Physics of Elementary Particles and Astrophysics\footnote{\url{https://astronu.jinr.ru/school/current}}.
This educational activity and participation in TAIGA and Baikal-GVD make ISU efficient and prospective for the training of the experts in the field.
\end{itemize}
Within these conditions we can effectively work on the challenges facing data and knowledge conservation, moreover we can test and evaluate our methods and approaches.
The neighborhood of the experiments measuring different messengers (TAIGA -- gamma and Baikal-GVD -- neutrino) and the large educational center (ISU) naturally lead one to the \textit{Baikal Multimessenger} concept, a testbed for the future multimessenger activity, which can be started within suggested Open Laboratory.

\section{The pillars of the Baikal Open Laboratory}
The main objectives of the future Open Laboratory are training experts and developing new instruments and methods for the multimessenger astronomy as well as supporting open software and open data initiatives.
Taking this and the present environment into account we can define the main pillars of it:

\begin{itemize}
\item \textit{Development open training programs}. All programs and their sources (i.e. scripts, slides, problems, etc.) developed in the frame of this Laboratory will be published online under free license and can be adopted by the third-party institutes and lecturers.
These lectures and seminars will be given at ISU (see below) and kept alive and updated.
\item \textit{Focus on modern IT and open source solutions}. The modern physics analysis suffers from the lack of the experts in big data and deep learning.
We plan to spend significant efforts on training of these experts during their education at ISU, attracting new experts and trying to save them in science.
Additionally we will focus on data analysis using modern methods~\cite{Bezyazeekov_DLC2019}.
\item \textit{Interaction between different facilities}.
The multimessenger astronomy implies data transfer between astroparticle experiments, 
which can be complicated by the data policies established by the different collaborations.
Within experiments located in the Baikal region we will focus on the policies, exchange protocols and software for the multimessenger astronomy.
\end{itemize}

\section{The current status of the development of Laboratory}
\label{sec:current_status}
For the time being there are few directions of the development of laboratory: \textit{online platform}, \textit{offline course} and \textit{training with operating experiments}.

\subsection{Online platform}
\begin{figure}
	\includegraphics[width=\textwidth]{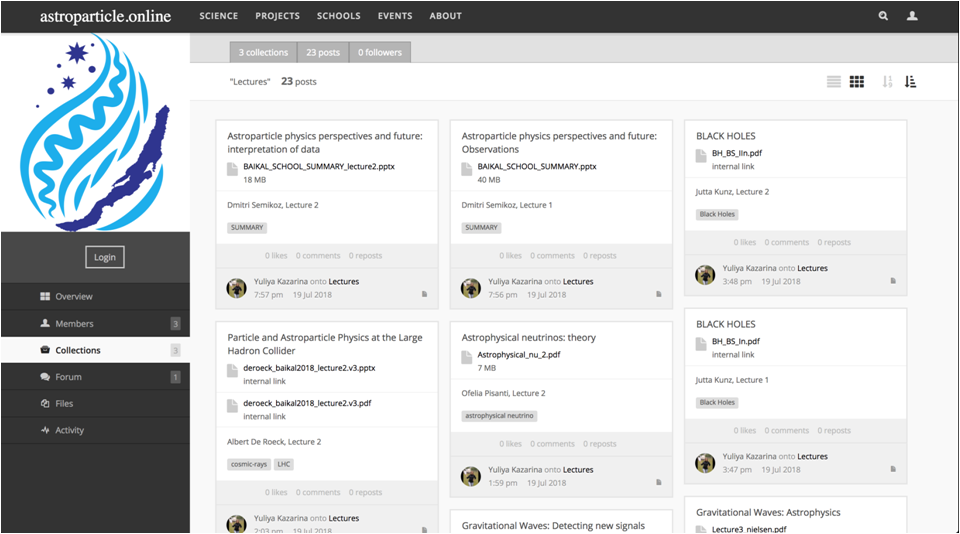}
	\caption{Screenshot of the Baikal-ISAPP summer school page on pilot version of \texttt{astroparticle.online}.
} \label{fig2}
\end{figure}

At the very beginning of GRADLCI we have established \texttt{astroparticle.online}, which aims at the following:
\begin{itemize}
\item \textit{Web-interface} for the open data services developed in the frame of GRADLCI. For details see Refs.~\cite{Shigarov_DLC2019,Nguyen_DLC2019,Shigarov:2018_GRID}.
\item \textit{Educational and outreach} materials in astroparticle physics, including VISPA-like interactive services~\cite{Bretz:2012}.
\item \textit{Enhancing the communication} between astrophysics.
We try to support networking by providing platform for partner experiments\footnote{see, e.g. \url{tunka-21cm.astroparticle.online} and \url{almarac.astroparticle.online}}, schools and events.
\end{itemize}
For the time being the portal is under construction, and the content is being filling.
We have successfully tested the pilot version (see Fig.~\ref{fig2}) of it at the ISAPP-Baikal Summer school\footnote{\url{https://astronu.jinr.ru/school/archive/school-2018}} as a collaboration framework. 
Moreover, the informational and interactive part of a new regular ISU course described below will be deployed within the portal.  
The servers of the platform have been deployed at the Matrosov Institute for System Dynamics and Control Theory\footnote{\url{http://idstu.irk.ru/en}}. 
At the very beginning we have used the open-source HUBzero platform~\cite{McLennan:2010}, 
however due to numerous technical problems and difficulties, it was decided to move to the widely used WordPress\footnote{\url{https://wordpress.com/}} and deploy all necessary plugins (e.g. VISPA) separately. 

\subsection{Offline educational course on astroparticle physics}
The Faculty of Physics of ISU has been established more than a century ago,
has a long history connected to physics and astrophysics, 
and many of the graduates of the Faculty work in the leading research institutes organizations around the world. 
Moreover, the students and graduates get an opportunity to work in TAIGA observatory and Baikal-GVD neutrino telescope during the study, 
many of them write bachelor and master theses in the frame of these experiments. 
Since ISU makes a great contribution to these experiments, we have decided to develop a new regular course \textit{``Introduction to experimental astroparticle physics''} for the bachelor and master students.
The course includes lectures, practical and laboratory works.

The theoretical part of the course consists of about ten lectures devoted to the ultra-high-energy cosmic rays, their origin and acceleration mechanisms, cosmic ray energy spectrum and its features, cosmic ray detection methods, gamma- and neutrino astronomy features, review of the largest astrophysical facilities.

The seminars of the course are focused on applied knowledge of simulations and data analysis.
We give an introduction to a major programming languages, namely \texttt{C}/\texttt{C++} and \texttt{Python}, to a main analysis frameworks, namely \texttt{ROOT}\footnote{\url{http://root.cern}}, \texttt{numpy}, \texttt{scipy} and \texttt{matplotlib}.
As a result of this course, students are able to solve problems in the modern data analysis.
The materials of the course are open-source and published online\footnote{\url{https://bitbucket.org/tunka/ap-seminar-latex/}}.

\begin{figure}[t]
	\includegraphics[width=1.0\textwidth]{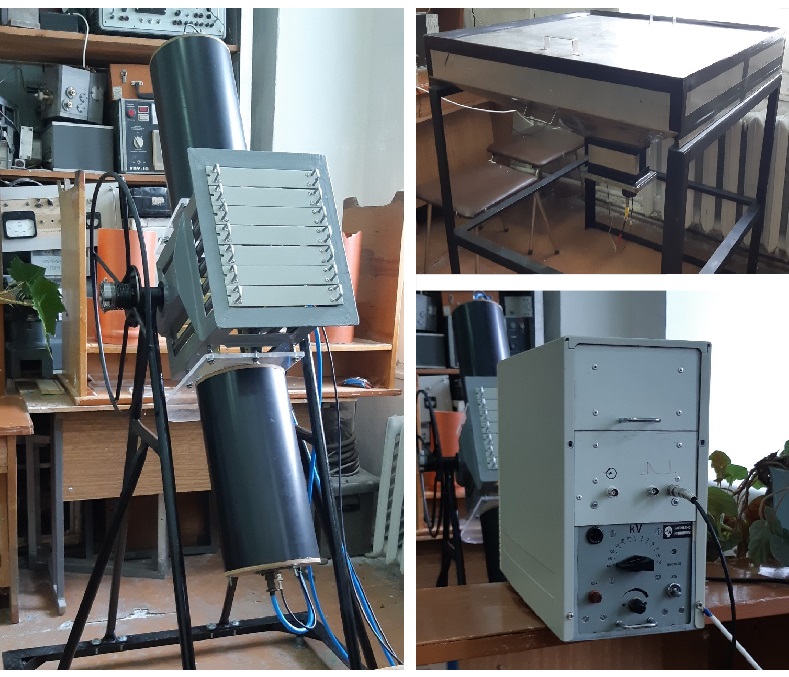}
	\caption{Laboratory setups developed for the astroparticle course.
	\textit{Left:} The telescope for studying the secondary component of cosmic rays.
	\textit{Top right:} The stand for studying the fluctuations of ionization losses.
	\textit{Bottom right:} The stand for study the main characteristics of photomultipliers (PMT).
	\label{fig1}
	}
\end{figure}

In the frame of this course three laboratory setups (see Fig.~\ref{fig1}) have been developed to familiarize students with the astrophysics detectors.
Students perform measurements on these setups, and analyze and interpret the data using knowledge obtained on the lectures and seminars.

As it was mentioned before, the materials and interactive part of the course will be incorporated in \texttt{astroparticle.online}.

\subsection{Training with operating experiments}
As was mentioned above, the students of ISU as well as visiting students have an opportunity to work with real hardware, software and data of TAIGA and Baikal-GVD.
We have an established workflow for the young scientists, which includes interview, training, data analysis and simulation, field works, etc.
This workflow has shown its efficiency, what has resulted in a number of significant results obtained by the young members of collaborations (see, e.g.~\cite{Marshalkina_ARENA2018,Shipilov_ARENA2018}).

\section{Conclusion}
After the years of the development of astrophysical experiments in the Baikal region we are ready to make a step further and establish an educational and outreach unit focused on multimessenger astronomy and open science.
Having unique environment ``on-site'' (ISU + TAIGA + Baikal-GVD), it is possible to develop and evaluate modern astrophysical methods and techniques very fast and efficient.
We have started from the open science activity declared in the frame of GRADLCI and are going to expand this to the Baikal Open Laboratory in Astroparticle Physics.
We hope that the future cooperation with GRADLCI will help us to share our ideas and progress to the global astroparticle community.


\section*{Acknowledgements}
This work was supported by Russian Science Foundation Grant 18-41-06003 
(Section \ref{sec:current_status}), by the Helmholtz Society Grant HRSF-0027 and by the Russian Federation Ministry of Education and Science (projects 14.593.21.0005 (Tunka shared core facilities, unique identificator RFMEFI59317X0005), 3.10131.2017/NM, 2017-14-595-0001-003, 3.9678.2017/8.9, 3.904.2017/4.6).
We are grateful to the members of the GRADLCI for the informational support of our activity.


%
%
%
%

\bibliographystyle{ieeetr}
\bibliography{references}

\begin{thebibliography}{10}

\bibitem{Kowalski_Bartos_mm}
I.~Bartos and M.~Kowalski, {\em Multimessenger Astronomy}.
\newblock 2399-2891, IOP Publishing, 2017.

\bibitem{Bychkov:2018zre}
I.~Bychkov {\em et~al.}, ``{Russian-German Astroparticle Data Life Cycle
  Initiative},'' {\em Data}, vol.~3, no.~4, 2018.

\bibitem{Budnev:2018fxf}
N.~Budnev {\em et~al.}, ``{TAIGA - a hybrid array for high energy gamma
  astronomy and cosmic ray physics},'' {\em EPJ Web Conf.}, vol.~191, p.~01007,
  2018.

\bibitem{Avrorin:2013uyc}
A.~D. Avrorin {\em et~al.}, ``{The prototyping/early construction phase of the
  BAIKAL-GVD project},'' {\em Nucl. Instrum. Meth.}, vol.~A742, pp.~82--88,
  2014.

\bibitem{Bezyazeekov_DLC2019}
P.~Bezyazeekov {\em et~al.}, ``{Advanced signal reconstruction in Tunka-Rex},''
  in {\em these proceedings}, 2019.

\bibitem{Shigarov_DLC2019}
I.~Bychkov {\em et~al.}, ``{Metadata extraction from raw astroparticle dataof
  TAIGA experiment},'' in {\em these proceedings}, 2019.

\bibitem{Nguyen_DLC2019}
M.~Nguyen {\em et~al.}, ``{Data aggregation in the Astroparticle Physics
  Distributed Data Storage},'' in {\em these proceedings}, 2019.

\bibitem{Shigarov:2018_GRID}
I.~Bychkov {\em et~al.}, ``{Using Binary File Format Description Languages for
  Documenting, Parsing, and Verifying Raw Data in TAIGA Experiment},'' {\em
  CEUR Workshop Proceedings}, pp.~563--567, 2018.

\bibitem{Bretz:2012}
H.~Bretz {\em et~al.}, ``{A Development Environment for Visual Physics
  Analysis},'' {\em JINST}, vol.~7, p.~T08005, 2012.

\bibitem{McLennan:2010}
M.~McLennan and R.~Kennell, ``{HUBzero: A Platform for Dissemination and
  Collaboration in Computational Science and Engineering},'' {\em Computing in
  Science and Engineering}, vol.~12, pp.~48--52, 2010.

\bibitem{Marshalkina_ARENA2018}
T.~Marshalkina {\em et~al.}, ``{First analysis of inclined air-showers detected
  by Tunka-Rex},'' vol.~ARENA2018 proceedings, 2018.

\bibitem{Shipilov_ARENA2018}
D.~Shipilov {\em et~al.}, ``{Signal recognition and background suppression by
  matched filters and neural networks for Tunka-Rex},'' vol.~ARENA2018
  proceedings, 2018.

\end{thebibliography}


\end{document}